\documentclass{article}
\usepackage[utf8]{inputenc}
\usepackage{siunitx}
\usepackage{graphicx}

\begin{document}

\title{HV discharges triggered by dual- and triple-frequency laser filaments}
\author{Thomas Produit,${}^{1}$ Pierre Walch,${}^{2}$ Guillaume Schimmel,${}^{1}$ Benoît Mahieu,${}^{2}$\\ Clemens Herkommer,${}^{3,4}$ Robert Jung,${}^{3}$ Thomas Metzger,${}^{3}$ Knut Michel,${}^{3}$\\Yves-Bernard André,${}^{5}$ André Mysyrowicz,${}^{6}$ Aurélien Houard,${}^{2}$\\ Jérôme Kasparian${}^{1,7,*}$ and Jean-Pierre Wolf${}^{1}$}
\date{${}^{1}$Groupe de Physique Appliquée, Université de Genève, Ch. de Pinchat 22, 1211 Geneva 4, Switzerland\\
${}^{2}$LOA, ENSTA ParisTech, CNRS, Ecole polytechnique, 828 Bd des Maréchaux, 91762 Palaiseau, France\\
${}^{3}$TRUMPF Scientific Lasers GmbH + Co. KG, Feringastr. 10a, 85774 Unterföhring, Germany\\
${}^{4}$Department of Physics, Technische Universität München, James-Franck-Str. 1, 85748 Garching, Germany\\
${}^{5}$LOA, Ecole polytechnique, CNRS, Route de Saclay, 91128 Palaiseau, France\\
${}^{6}$André Mysyrowicz Consultants, 6 Rue Gabriel, 78000 Versailles, France\\
${}^{7}$Institute for Environmental Sciences, Université de Genève, Bd Carl Vogt 66, 1211 Geneva 4, Switzerland\\
${}^{*}$jerome.kasparian@unige.ch
}

\maketitle

\begin{abstract}
We study the use of frequency upconversion schemes of near-IR picosecond laser pulses and compare their ability to guide and trigger electric discharges through filamentation in air. Upconversion, such as Second Harmonic Generation, is favorable for triggering electric discharges for given amount of available laser energy, even taking into account the losses inherent to frequency conversion.
We focus on the practical question of optimizing the use of energy from a given available laser system and the potential advantage to use frequency conversion schemes.
\end{abstract}

\section{Introduction}

The first attempts to control and trigger high-voltage discharges and lightning date back to the 1970‘s. They were based on high power and high-energy Nd:Glass laser or CO${}_2$ nanosecond laser pulses and proved to be prohibitive for a distant operation \cite{Vaill1970, Koopman1971, Saum1972, Miki1993, Shindo1993, Shindo1993bis, Wang1994, Wang1995, Miki1996, Uchida1999, Wolf2018}. Indeed, they required pulse energies of almost hundred Joules and their focusing resulted in avalanche ionization at the focus, with electron densities up to \SI{e20}{\per\centi\meter\cubed}, above the critical plasma density in the mid-IR. This hindered therefore further laser propagation as the leading edge of the beam rendered the air opaque to its trailing edge. \cite{Wolf2018} offers a full review on the subject.

The advent of ultrashort (ps and shorter), high intensity (in the \si[per-mode=repeated-symbol]{\tera\watt\per\centi\meter\squared} range) laser pulses based on chirped-pulse amplification \cite{Strickland1985} led to the observation of filamentation in air \cite{Braun1995, Couairon2007, Berge2007, Chin2005}. Filamentation is a dynamically self-guided propagation regime in which self-focusing by the Kerr effect and defocusing by the generated plasma and other higher-order effects, result in long, narrow ionized channels with electron densities of \SI{e15}{\per\centi\meter\cubed} - \SI{e16}{\per\centi\meter\cubed}. Filaments are therefore conducting, and were early proposed for guiding high-voltage discharges and lightning \cite{Diels1992, Zhao1995}. The first experiments have been performed in the UV \cite{Diels1992, Zhao1995, Rambo1999}, where ionization is more efficient due to the higher photon energy. However, a limitation to spatially extend the electric discharge guiding by laser filaments is the lifetime of the plasma created by the laser \cite{Wolf2018}. On a timescale of \SI{1}{\micro\second} the carrier  density decreases by about four orders of magnitude \cite{Wolf2018}, corresponding to only about a meter of propagation at a typical laser-guided leader speed of \SI[per-mode=repeated-symbol]{e5}{\meter\per\second} - \SI[per-mode=repeated-symbol]{e6}{\meter\per\second} \cite{Wolf2018, Pepin2001, Rodriguez2002}.

Efforts to overcome that limitation focused on optimizing the wavelength, from the UV \cite{Diels1992, Zhao1995, Zvorykin2010, Ionin2012}, the visible \cite{Liu2011}, the near-IR \cite{Rambo1999, Pepin2001, Rodriguez2002, Liu2011, Fujii2008, Forestier2012}, up to \SI{4}{\micro\meter} in the Mid-IR \cite{Mongin2016}.

Longer wavelengths, while less efficient in ionizing the air, are prone to produce longer and higher energy baring single filaments. Another approach consists in re-detaching and/or re-heating the free electrons by subsequent laser pulses. On one hand, multiple ultrashort pulses \cite{Schubert2016, Zhang2009, Lu2015} increased the free electron lifetime over a duration similar to the pulse train.

On the other hand, dual-frequency fs-ns sequences \cite{Mejean2006, Schubert2017, Papeer2014} were investigated, with mitigated results ranging from a clear decay of the breakdown threshold by 5\% \cite{Mejean2006}, to an acceleration of the discharges by a factor of two without a significant increase of the discharge probability \cite{Schubert2017}. Finally, Scheller et al. \cite{Scheller2014} report a surprising reduction of the breakdown field up to a factor of 10 for electrode separations up to \SI{25}{\centi\meter}, with the help of a ns-heater laser and a focusing axicon, but at the expenses of lineic energies deposition as large as \SI[per-mode=repeated-symbol]{1}{\joule\per\centi\meter}.

Here we combine these multi-pulse and multi-frequency approaches by investigating simultaneous dual- or triple-frequency pulses combining the fundamental and second and/or third harmonics of a Yb:YAG ultrashort laser. Recently, it was shown that multiple-frequency multi-photon pathways can significantly contribute the ionization efficiency~\cite{Doussot2017,Yuan2018,Tian2018,Neufeld2018}.

We compare the discharge triggering efficiency of various frequency combinations, as well as the associated time delay with the same laser pulse without frequency conversion.
We therefore focus on the practical question of optimizing the use of energy from a given available laser system and the potential advantage to use frequency conversion schemes. We show that indeed frequency doubling and tripling schemes favors discharge triggering at the centimeter scale despite their intrinsic non-unity conversion efficiency.

\begin{figure}[t]
\centering\includegraphics[width=0.9\columnwidth]{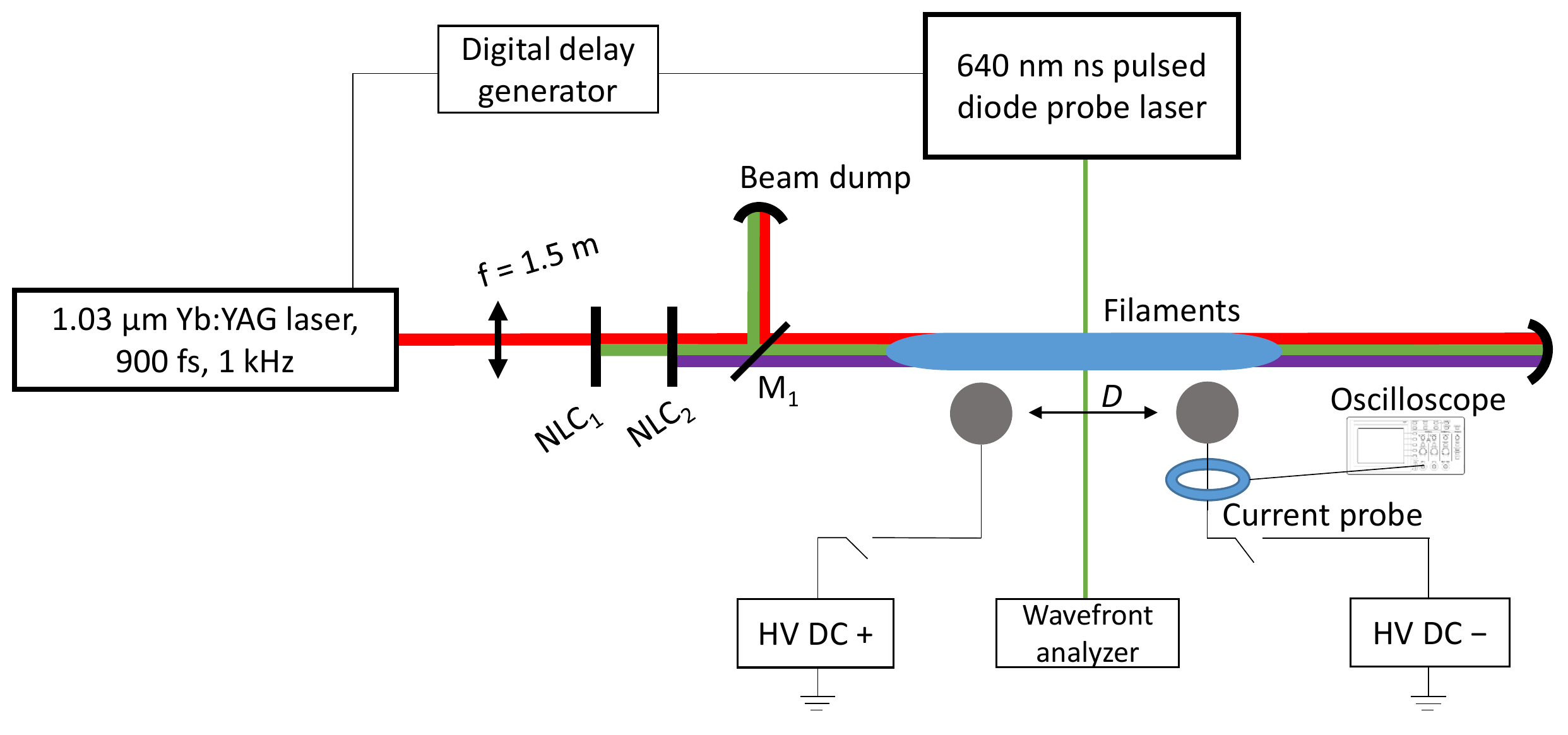}
\caption{Experimental setup. NLC${}_{1,2}$ are LBO-crystal for second and third harmonic generation, respectively. M${}_1$ is an optional set of mirrors to filter the different harmonics. HV DC +/- are respectively the positive and the negative electrodes of the high voltage (HV) generator, which are separated by the distance $D$.}
\label{fig:setup}
\end{figure}

\section{Experimental setup}

The experimental setup is sketched in \mbox{Fig. \ref{fig:setup}}. Two spherical electrodes of diameter $d$ = \SI{1.2}{\cm} were placed at a distance up to $D$ = \SI{10}{\mm} from each other. Both electrodes were connected to a direct current (DC) high voltage (HV) generator supplying adjustable voltages of opposite polarities up to +/– \SI{13}{\kilo\volt}, i.e. a maximum voltage $V$ = \SI{26}{\kilo\volt} between the electrodes. The temperature and relative humidity of the experimental environment were stabilized to T = \SI[separate-uncertainty]{23.0(5)}{\celsius} and RH = \SI[separate-uncertainty]{35(10)}{\percent} throughout the whole experiments, monitored automatically in the experimental hall.

A Yb:YAG thin disk, chirped pulse amplification (CPA) laser system (Dira 200-1 TRUMPF Scientific Lasers GmbH + Co. KG \cite{Klingebiel2015}) delivered pulses with energies up to \SI{165}{\milli\joule} and \SI{900}{\fs} duration at \SI{1030}{\nm} at a repetition rate of \SI{1}{\kHz}. The beam with a diameter of \SI{17}{\mm}, was slightly focused by an $f$ = \SI{1.5}{\m} lens. 
It produced \SI{75}{\cm}-long filaments (covering the whole gap between the electrodes), at a transverse distance $L$ = \SI{1}{\mm} from the electrodes. Optionally, non-linear LBO-crystals were inserted in the beam path to generate second (NLC${}_1$, LBO-crystal for Type I SHG, \SI{33 x 33 x 1.5}{\mm}) and/or third harmonics (NLC${}_2$, LBO-crystal for Type II THG, \SI{10 x 10 x 3}{\mm}) respectively.

\begin{figure}[t]
\centering\includegraphics[width=0.6\columnwidth, keepaspectratio]{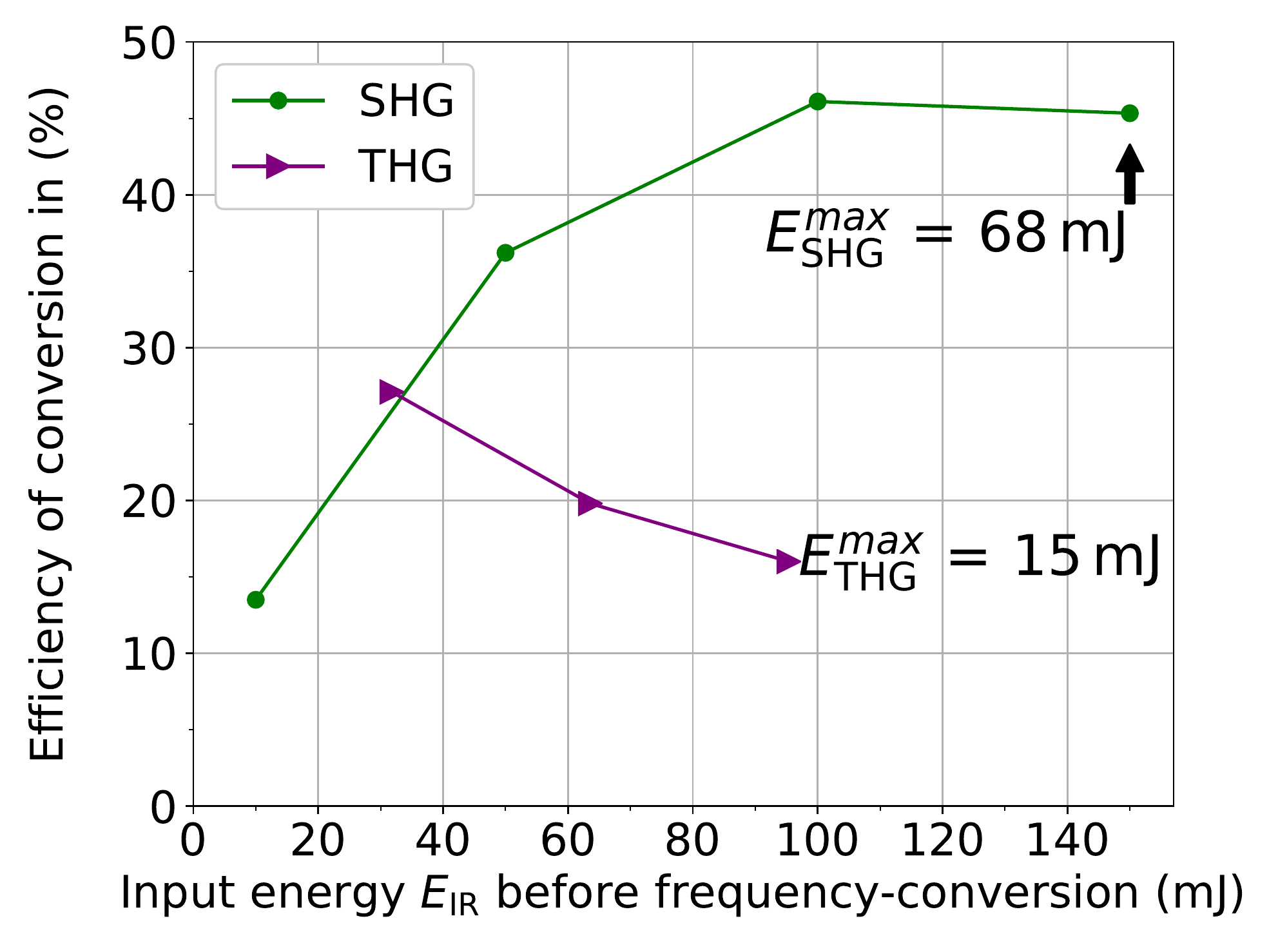}
\caption{Energy conversion efficiency $\eta_{\mathrm{upconverted}} = {E_{\mathrm{upconverted}}}/{E_{\mathrm{IR}}}$ as a function of the input energy $E_{\mathrm{IR}}$. }
\label{fig:efficiency}
\end{figure}

The energy conversion efficiencies reached up to $\eta_{\mathrm{SHG}} = \frac{E_{\mathrm{SHG}}}{E_{\mathrm{IR}}} = $ \SI{45}{\percent} and $\eta_{\mathrm{THG}} = \frac{E_{\mathrm{THG}}}{E_{\mathrm{IR}}} = $ \SI{27}{\percent} for the second and third harmonic generation, respectively, as can be seen on \mbox{Fig. \ref{fig:efficiency}}. This allowed delivering \SI{68}{\milli\joule} at \SI{515}{\nm} (along with \SI{83}{\milli\joule} of residual \SI{1030}{\nm}) and \SI{16}{\milli\joule} at \SI{343}{\nm} (along with a few \si{\milli\joule} at \SI{515}{\nm} and \SI{130}{\milli\joule} of residual \SI{1030}{\nm}). The SHG conversion efficiency saturates at $\eta_{\mathrm{SHG}}$ = \SI{45}{\percent} for input energies $E_{\mathrm{IR}}$ above \SI{100}{\milli\joule} (see \mbox{Fig. \ref{fig:efficiency}}). In contrast, the THG conversion efficiency decays for increasing laser energy $E_{\mathrm{IR}}$. This indicates that the NLC${}_2$ crystals longitudinal length of \SI{3}{\mm} is too long and a back conversion occurs. Furthermore, transverse dimensions of the frequency-tripling crystal NLC${}_2$ (\SI{10 x 10 x 3}{\mm}) clipped the laser beam and hence prevented to take advantage of the full beam energy. Nonetheless, energies up to \SI{16}{\milli\joule} at \SI{343}{\nm} were recorded. An optional set of dichroic mirrors (M${}_1$ on \mbox{Fig. \ref{fig:setup}}) allowed filtering out the harmonics and/or the residual \SI{1030}{\nm} from the beam.

To assess the repetition rate effect on electric discharges, the results obtained with the thin disk Yb:YAG laser were compared to those of a Ti:Sa CPA system. This system delivered up to \SI{300}{\milli\joule} at \SI{800}{\nm} at a repetition rate of \SI{10}{\hertz} (ENSTAMobile, Amplitude TT mobile). Its \SI{50}{\femto\second} pulses were chirped to \SI{1}{\pico\second}, close to the pulse duration of the Yb:YAG laser and the optical geometry was chosen to match the numerical aperture of the previous conditions and hence to provide a consistent comparison scheme. In each experimental condition, the breakdown threshold was determined systematically by slowly and continuously decreasing the voltage on the electrodes until discharges stopped to occur. This threshold voltage was then converted into an average electric field, considering a homogeneous electric field between the electrodes over the full inter-electrode distance $D$. Similarly, for a voltage fixed to its maximum value ($V$ = \SI{26}{\kilo\volt}), we determined the maximum discharge length in given conditions by increasing the inter-electrode distance $D$ in sub-\si{\mm} steps until no more discharge are recorded. Finally, the breakdown time delays between the breakdown current peak and the laser shot were recorded using the current probe traces on the oscilloscope.

The temporal dynamics of the discharge was characterized via the probability distribution function (PDF) of the time delays between the laser pulse and the discharge. This function can be seen as a normalized histogram of the delays. We compared the decay of the discharge probability as a function of delay with that of the free electrons, as simulated by the model described in \cite{Schubert2016}. This kinetic model implements multiphoton/tunnel ionization, electron attachment to neutral molecules, photodetachment, ion-ion and electron-ion recombination, impact ionization and excitation.

It considers the partition of the thermal energy between the kinetic energy of the electrons and the heavy species, as well as the molecular vibrational energy. It disregards spatial effects like hydrodynamics, transport, as well as electric field inhomogeneities and screening.

In order to better characterize the local effect of filaments on the air density between the electrodes, we transversely mapped the dephasings induced by the beam \SI{1}{\micro\second} – \SI{900}{\micro\second} after the pulses, in an interferometric setup detailed in \cite{Point2014, Point2015} and based on a pulsed diode laser source at \SI{640}{\nm} and a Phasics SID4 HR wavefront analyzer.

\section{Results}

\subsection{Electric field and distance breakdown thresholds}

Figure 3(a) compares the electric field breakdown threshold for various frequency combinations, as a function of the energy delivered by the laser before the frequency conversion (input energy $E_{\mathrm{IR}}$), for a distance up to $D$ = \SI{10}{\mm} between the electrodes.
Above several tens of \si{\milli\joule}, switching on the laser at \SI{1030}{\nm} decreases the breakdown voltage by \SI{30}{\percent} to \SI{50}{\percent} of the natural breakdown field $E_{\mathrm{nat}}$ = \SI[per-mode=repeated-symbol]{28.6}{\kilo\volt\per\cm}, consistent with previous results \cite{Diels1992, Pepin2001, Rodriguez2002, Houard2016}. The breakdown voltage progressively decreases with increasing laser pulse energy then stabilizes and even tends to rise again when the input laser energy is further increased. 
Successively adding the LBO crystals generating the second and third harmonics reduces the laser energy at which the minimum of the breakdown voltage is reached, evidencing an increase of the efficiency of the discharge triggering by the laser.

In the case of chirped pulses at \SI{800}{\nm} at a repetition rate of \SI{10}{\hertz} (black curve), the breakdown electrical field stabilizes at V \textasciitilde  \SI[per-mode=repeated-symbol]{17}{\kilo\volt\per\cm} for input energies above \SI{50}{\milli\joule}, without re-increasing at higher pulse energies.

\begin{figure}
	\centering\includegraphics[width=9cm]{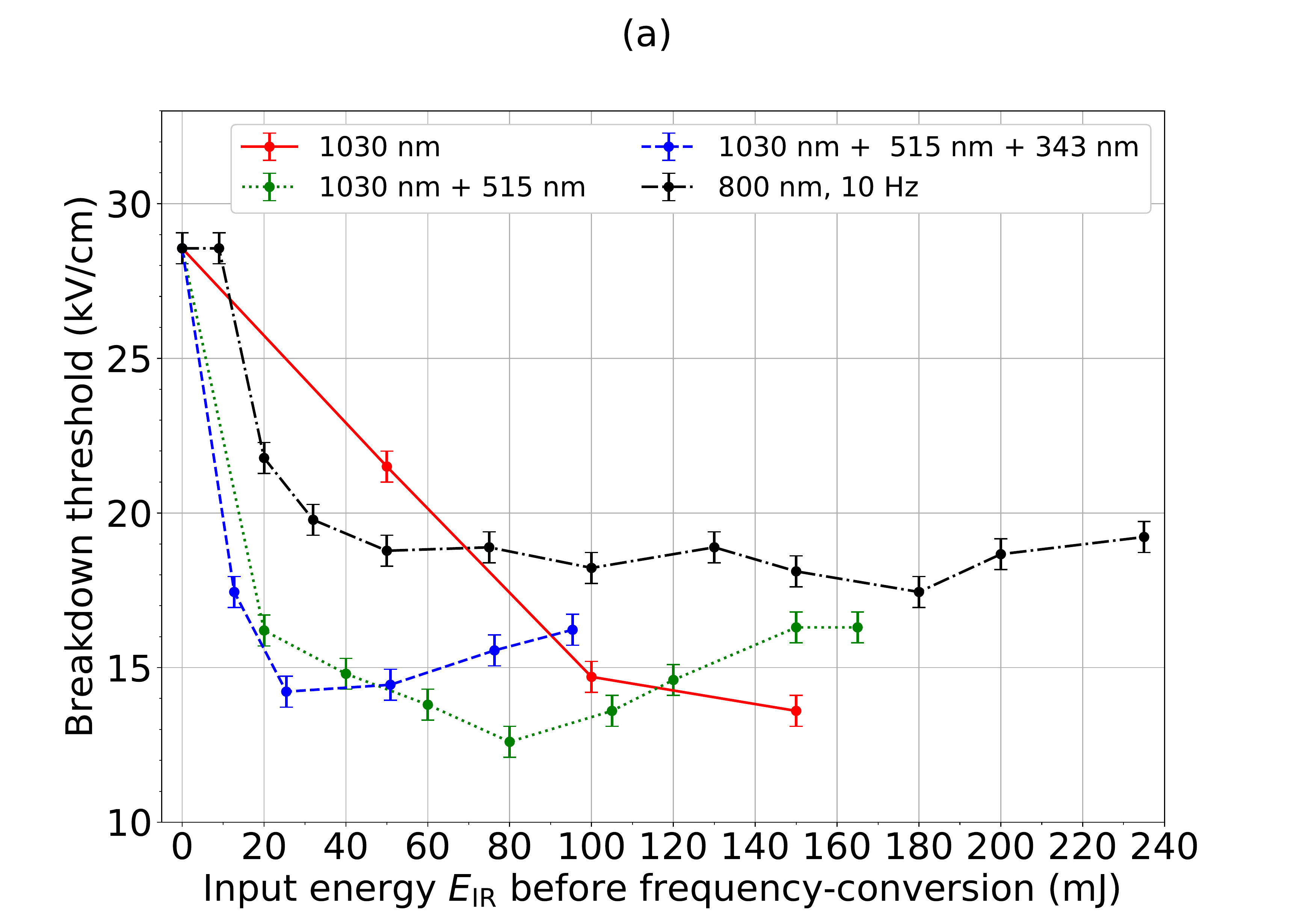} 
	\\
	\centering\includegraphics[width=9cm]{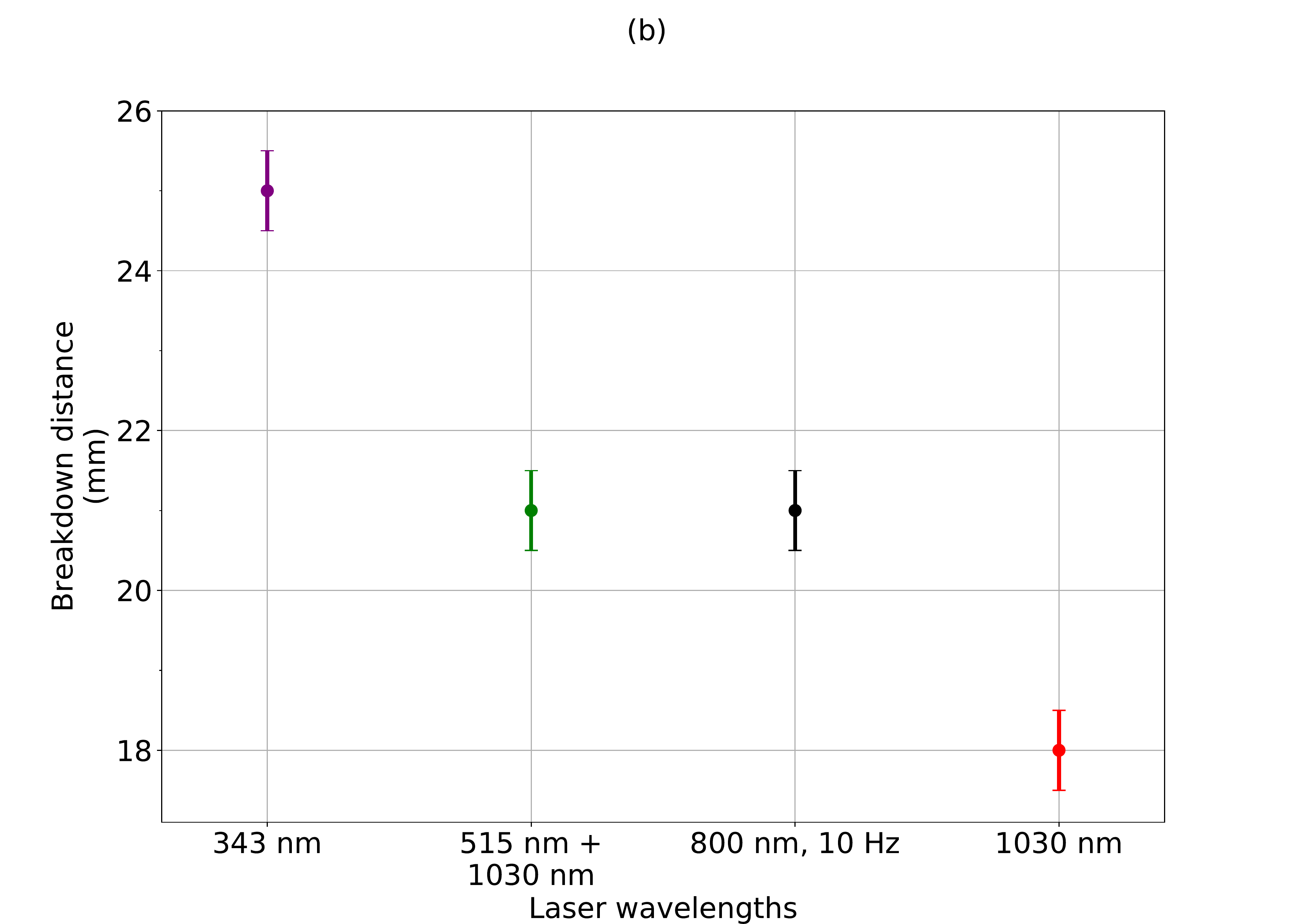}
	\caption{(a) Breakdown threshold as a function of the input laser energy.  (b) Maximum breakdown distance for the different harmonics and the 800 nm at 10 Hz. Except for the 343 nm case, the input energy is 150 mJ; at 343 nm, input energy is 100 mJ. The error bars indicate the systematic error from the readout procedure, which in both graphs is the dominating error source.}
	\label{fig:breakdown}
\end{figure}

Consistent with the lower breakdown electric field, the breakdown distance for an increased input energy is longer for shorter wavelengths (up to \SI{25}{\mm} at \SI{343}{\nm}, as compared to \SI{18}{\mm} at \SI{1030}{\nm}), again evidencing a more efficient discharge triggering (see Fig.~\ref{fig:breakdown}(b)).
It is remarkable that this longer breakdown distance is obtained in spite of the \SI{25}{\percent} energy conversion efficiency into the UV, so that frequency upconversion of a mid-IR laser beam, even after filtering away the remaining fundamental and second harmonic, actually improves discharges triggering.

\subsection{Breakdown time delay}

\begin{figure}
	\centering\includegraphics[width=\columnwidth]{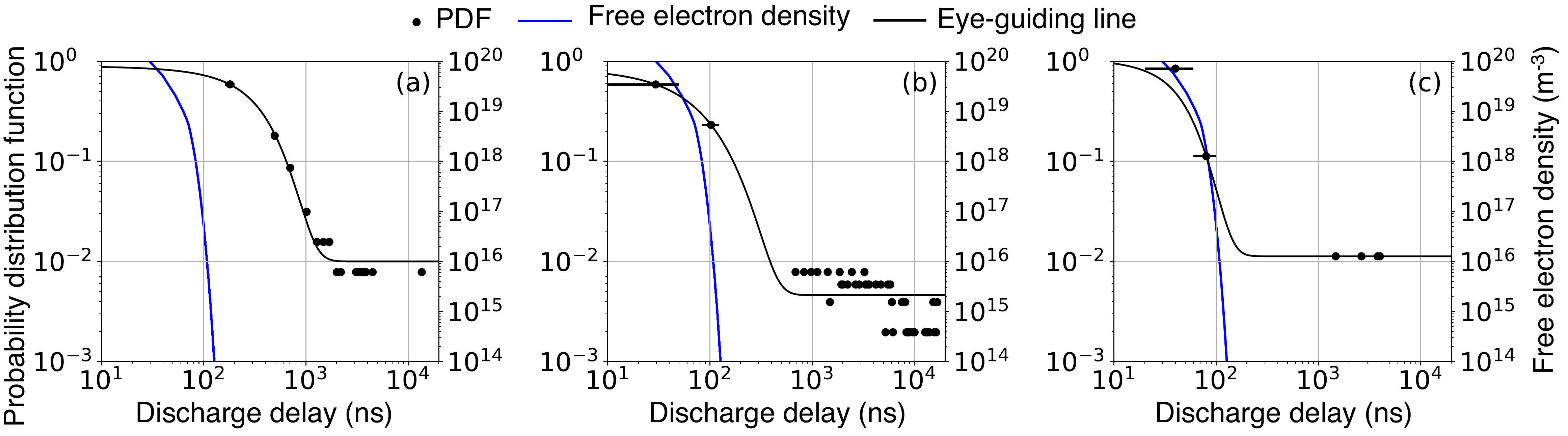}
	\caption{Probability distribution function of the discharge delay after the laser pulse. The black solid line serves as eye guiding feature. The temporal evolution of the electron simulated with the model described in \cite{Schubert2016} is shown in blue. (a) Input laser energy of \SI{50}{\milli\joule} at \SI{1030}{\nm} (128 discharges in total). (b) Input laser energy of \SI{150}{\milli\joule} at \SI{1030}{\nm} (512 discharges in total). (c) Input laser energy of \SI{28}{\milli\joule} at \SI{1030}{\nm} + \SI{12}{\milli\joule} at \SI{515}{\nm} (88 electric discharges in total).}
	\label{fig:delay}
\end{figure}

Besides the breakdown threshold, the laser strongly influences the the discharge occurrence probability, as well as the temporal dynamics of discharges characterized by the PDF of the delay between the laser pulse and the discharge. 
For energies below the optimum (\SI[per-mode=repeated-symbol]{22}{\kilo\volt\per\cm}, \SI{< 50}{\milli\joule} at \SI{1030}{\nm}, \mbox{Fig. \ref{fig:delay}(a)}), the discharge occurrence probability is low, only \SI{18}{\percent} of the laser pulses induce an electric discharge. Although all discharges are fully guided along the laser path, their establishment occurs at least \SI{180}{\ns} and at most \SI{5}{\micro\second} after the laser pulse, \SI{85}{\percent} of them occurring within \SI{700}{\ns} from the laser pulse.

At the optimum, (\textasciitilde \SI[per-mode=repeated-symbol]{13}{\kilo\volt\per\cm}, reached either for \SI{150}{\milli\joule} at \SI{1030}{\nm}, \mbox{Fig. \ref{fig:delay}(b)}, or for \SI{28}{\milli\joule} at \SI{1030}{\nm} along with \SI{12}{\milli\joule} at \SI{515}{\nm}, \mbox{Fig. \ref{fig:delay}(c)}) almost each laser pulse triggers a guided spark, even at \SI{1}{\kilo\hertz} repetition rate. The discharge occurrence probability is \SI{92}{\percent} for the \SI{150}{\milli\joule} at \SI{1030}{\nm} and \SI{88}{\percent} for the dual-frequency case (\SI{28}{\milli\joule} at \SI{1030}{\nm} along with \SI{12}{\milli\joule} at \SI{515}{\nm}). In both cases, this high discharge occurrence probability is accompanied by a fast discharge triggering: \SI{81}{\percent} of the discharges occur within \SI{100}{\ns} after a laser pulse, and all of them within \SI{2}{\micro\second} (\mbox{Fig. \ref{fig:delay}(b)}). In the dual-frequency case, \SI{95}{\percent} of the discharges even occur within \SI{80}{\ns} after a laser pulse, and all of them within \SI{4}{\micro\second} (\mbox{Fig. \ref{fig:delay}(c)}). 

\subsection{Phase map and low-density channel}

With the help of the transverse interferometric setup (see \mbox{Fig. \ref{fig:setup}}), we measured the low-density channel created in the wake of the filament. The strong heating of the filament region due to successive filaments at the laser repetition rate of \SI{1}{\kilo\hertz} results in an asymmetric pattern of dephasing with a higher density above the filaments and a lower density below. This pattern remains stable between each laser shot. This phase shift is representative of the air density in different regions around the filament. We performed a semi-quantitative analysis comparing the regions respectively \SI{11}{\mm} above and \SI{11}{\mm} below the filament. 

\begin{figure}
	\centering\includegraphics[width=0.6\columnwidth, keepaspectratio]{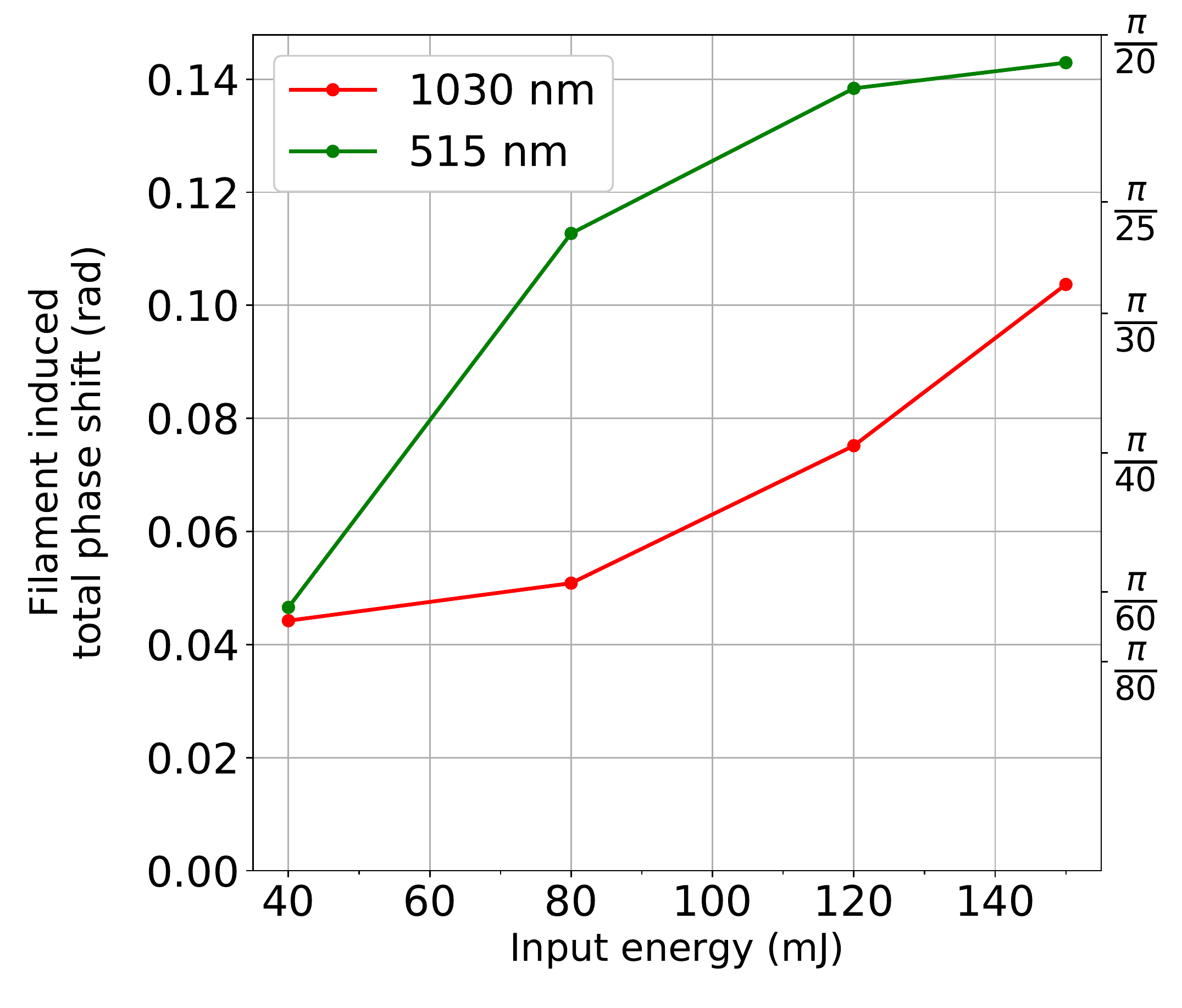}
	\caption{Depth of the phase shift induced in the wake of the filaments measured \SI{1}{\micro\second} after the laser pulse.}
	\label{fig:phase}
\end{figure}

The phase shift is larger between these regions (see \mbox{Fig. \ref{fig:phase}}) at all laser input energies for the frequency doubled pulse alone (\SI{515}{\nm}, \SI{1030}{\nm} filtered away) as compared to the fundamental. This deeper phase shift indicates a deeper low-density channel for this wavelength, thus a better discharge guiding ability. For instance at \SI{80}{\milli\joule} laser energy, the phase shift for the \SI{515}{\nm} exceeds the one of the fundamental (\SI{1030}{\nm}) by a factor of two (see \mbox{Fig. \ref{fig:phase}}). This maximum discrepancy coincides with the minimum of the electric field breakdown threshold, i.e the optimal discharge triggering pulse energy for the frequency-doubled scheme (see \mbox{Fig. \ref{fig:breakdown}}).

\section{Discussion and conclusion}

Our experiments allow to identify three regimes of laser-triggered DC discharges.

At low pulse energy (\SI{50}{\milli\joule} at \SI{1030}{\nm}), the breakdown voltage reduction is small (about \SI{33}{\percent}), and the laser is not sufficiently strong to systematically trigger discharges, i.e. the discharge probability is low (\textasciitilde\SI{18}{\percent}). Furthermore the discharge build-up time is long (\SI{\ge 180}{\ns}), exceeding the lifetime of the electrons (see blue curve in \mbox{Fig. \ref{fig:delay}(a)}).

At higher ionization rates (\SI{>100}{\milli\joule} at \SI{1030}{\nm}, \SI{>80}{\milli\joule} input energy with SHG conversion or \SI{25}{\milli\joule} converted to the THG) that also correspond to higher energy deposition, the electron density is sufficient to almost systematically trigger an avalanche. This occurs within the \textasciitilde\SI{100}{\ns} lifetime of the electrons, as illustrated by the blue curve in Fig.~\ref{fig:delay}(b). In addition to that, the more pronounced air density depletion coming from the higher energy deposition implies a faster decrease of the breakdown voltage described by Paschen’s law \cite{Paschen1889, Tzortzakis2001}.

The dephasing between regions around filaments (see \mbox{Fig. \ref{fig:phase}}) shows indeed a more pronounced density depletion for laser condition corresponding to higher ionization rates. This observation is consistent with previous experimental and numerical evidence that the air density depression in the wake of the filamentation has a time scale of several ms \cite{Wolf2018, Jhajj2014, Point2015}, longer than the time between the laser shots. Both the air density depression and the high electron density explain satisfactorily the reduced discharge build-up time compared to those in the low ionization regime. In these optimal conditions, the breakdown voltage is moreover reduced by more than a factor two (\mbox{Fig. \ref{fig:breakdown}}).

Even further increasing the laser energy and/or the ionization efficiency (with frequency-converting schemes) does not result in more efficient discharge triggering as expected, but rather in a decay of the apparent laser effect. In this regime, one cannot neglect the effect of filaments on the HV generator itself. Indeed, as reported in \cite{Schubert2015}, the filaments spread free charges between the electrodes, allowing a diffuse current flow unloading the generator and ultimately leading to electrical arc suppression, in a process similar to ultra-corona \cite{Rizk2010}. A sufficient permanent concentration of free charges guarantees a remaining conductivity that unloads the generator, reducing its effective voltage, and hence the effective electric field between the electrodes.

In contrast at a lower repetition rate, neither the air density depletion nor the free charge carrier density are maintained between pulses, evidencing the specific dynamics at the \si{\kHz} repetition rates. 

As expected from \cite{Liu2011} frequency-doubled filaments are more efficient in lowering the breakdown threshold than their fundamental counterpart. Furthermore, the breakdown voltage observed with  the triple-frequency scheme (blue curve on \mbox{Fig. \ref{fig:breakdown}}) reaches a minimum of \SI[per-mode=repeated-symbol]{14}{\kilo\volt\per\cm} already at \SI{25}{\milli\joule} laser input energy. Moreover, the breakdown distance is maximized by filaments at \SI{343}{\nm} also indicating a more homogeneous lineic plasma distribution. This higher efficiency of shorter wavelength pulses obviously stems from the higher ionization rates related to the higher photon energy. More ionization simultaneously implies more free charges and more energy deposition, hence a lower air density, and finally less efficient attachment to neutral molecules and therefore a longer plasma lifetime.

The plasma evolution model, described in \cite{Schubert2016} confirms this interpretation. Indeed, simulations show that for picosecond pulses at \SI{1030}{\nm} the tunnel ionization term widely dominates over the multiphoton ionization, described by the Perelomov-Popov-Terent'ev (PPT) model. In this ps regime the laser also deposits more heat than in the fs regime. The second and third harmonics (at \SI{515}{\nm} and \SI{343}{\nm}, respectively) on the other hand yield higher PPT ionization yields than the fundamental, due to the larger energy per photon. Furthermore, they produce more filaments due to their lower critical power (scaling with $\lambda^2$), even when taking into account the conversion efficiencies.

Hence, a synergy between frequencies can occur in the dual- or triple-frequency schemes, via two simultaneous processes. First, the multi-photon, multi-frequency can increase the ionization yield~\cite{Doussot2017,Yuan2018,Tian2018,Neufeld2018}. Second, the increased ionization induced by the shorter wavelength provides more seeds for avalanche ionization driven predominantly by the fundamental wavelength at \SI{1030}{\nm} and an increased energy deposition. The resulting increased ionization rate for the dual- or triple-frequency case will therefore contribute positively to the electric discharge triggering. This increased efficiency may even be combined with the self-healing capability of Bessel beams in order to circumvent obstacles, ad demonstrated by Clerici et al.~\cite{Clerici2015}.

In conclusion, we showed that, given a fixed amount of laser energy, upconverting the laser light and creating dual- and even triple-frequency filaments widely increases the efficiency of electric discharge triggering.

\section*{Funding}

Horizon 2020 Framework Programme (H2020) (737033-LLR)

\section*{Acknowledgments}

We  gratefully  acknowledge technical support from Michel Moret.


\end{document}